\documentclass[11pt,letterpaper]{article}

\usepackage[letterpaper,margin=1.0in]{geometry}
\usepackage{cite}
\usepackage{amsmath,amssymb,amsfonts}
\usepackage{algorithmic}
\usepackage{graphicx}
\usepackage{textcomp}
\usepackage{xcolor}
\usepackage{url}
\usepackage{amssymb}
\usepackage{amsthm}
\usepackage{multirow}
\usepackage{algorithm}
\usepackage{color}
\usepackage{framed}

\usepackage[pdftex,
            colorlinks,
            plainpages=false,
            hyperindex,
            bookmarksopen,
            linkcolor=black,
            citecolor=black,
            urlcolor=blue]{hyperref}

\usepackage{authblk}
\usepackage[normalem]{ulem}

\newcommand{\ket}[1]{|{#1}\rangle}

\def\BibTeX{{\rm B\kern-.05em{\sc i\kern-.025em b}\kern-.08em
    T\kern-.1667em\lower.7ex\hbox{E}\kern-.125emX}}
\begin{document}

\title{Private Set Intersection with Delegated Blind Quantum Computing}

\author[1,2]{Michele Amoretti\footnote{michele.amoretti@unipr.it}}
\affil[1]{Department of Engineering and Architecture - University of Parma, Italy}
\affil[2]{Quantum Information Science @ University of Parma, Italy}

\maketitle

\begin{abstract}
Private set intersection is an important problem with implications in many areas, ranging from remote diagnostics to private contact discovery. In this work, we consider the case of two-party PSI in the honest-but-curious setting. We propose a protocol that solves the server-aided PSI problem using delegated blind quantum computing. More specifically, the proposed protocol allows Alice and Bob (who do not have any quantum computational resources or quantum memory) to interact with Steve (who has a quantum computer) in order for Alice and Bob to obtain set intersection such that privacy is preserved. In particular, Steve learns nothing about the clients' input, output, or desired computation. The proposed protocol is correct, secure and blind against a malicious server, and characterized by a quantum communication complexity that is linear in the input size.

\textbf{keywords} - \textit{private set intersection, measurement-based quantum computing, delegated blind quantum computing}
\end{abstract}

\section{Introduction}

\textit{Private set intersection (PSI)} is a problem within the field of secure computation. 
In two-party PSI, Alice and Bob each hold a set of $m$ items, i.e., $\mathcal{A} = \{a_1,..,a_m\}$ and $\mathcal{B} = \{b_1,..,b_m\}$, respectively.
The goal is to design a protocol by which Alice and Bob obtain the intersection $\mathcal{A} \cap \mathcal{B}$, under the privacy restriction that anything about items that are not in the intersection must not be revealed. I.e., if $b_i \notin \mathcal{A}$ then Alice learns nothing about it.

PSI is an important problem with implications in many areas. For example, in \emph{remote diagnostics} \cite{Brickell2007}, a vectorized patient's (client) electronic health record gets a status (sick or not sick with a certain disease) from a medical diagnostic program. While the client learns about her sickness, the program remains secret and the program owner (server) does not learn anything about the client's data. Another example is \emph{private record linkage} \cite{He2017}, where two data owners hold different types of information for the same customer. In order to make data mining possible, the two records must be linked together and shared without giving away any other private data stored. In \emph{private contact discovery}, a user (client) wants to find out who of its private contacts also have a certain communication app (server) \cite{Demmler2018}. In \emph{DNA testing and pattern matching} \cite{Yanai2020}, the user gets its DNA sequenced and wants to find out about sequences linked to genetic diseases which are stored on a database (server). 

\begin{figure*}[ht!]
	\centering
	\includegraphics[width=10cm]{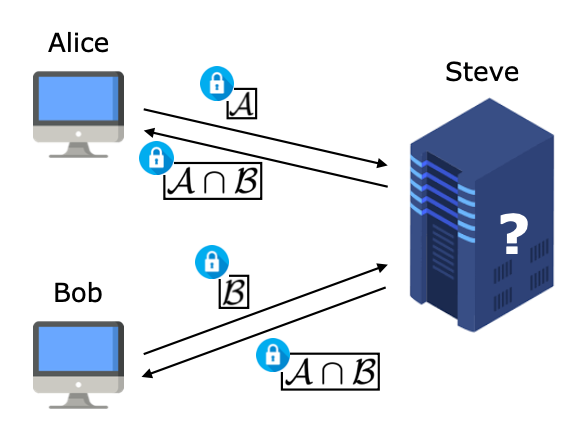}
	\caption{Server-aided PSI.}
	\label{fig:SA-PSI}
\end{figure*}

In this work, we consider the case of two-party PSI in the honest-but-curious setting. I.e., Alice and Bob fairly cooperate to obtain $\mathcal{A} \cap \mathcal{B}$, but they would love to know the other party's full set. We assume that Alice and Bob agree to use the service of an untrusted third party. In this setting, also known as \emph{server-aided PSI}, Alice and Bob interact with an additional party, called Steve, under the following privacy restriction: Steve should not learn information about the items of Alice and Bob (Figure \ref{fig:SA-PSI}). We also assume neither Alice nor Bob collude with Steve to break the other party's privacy.

We use delegated blind quantum computing, in the multiparty version introduced by Kashefi and Pappa \cite{Kashefi2017}, to solve  the server-aided PSI problem.
More precisely, we give a protocol that allows Alice and Bob (who do not have any quantum computational resources or quantum memory) to interact with Steve (who has a quantum computer) in order for Alice and Bob to obtain $\mathcal{A} \cap \mathcal{B}$ such that privacy is preserved. In particular, Steve learns nothing about Alice's and Bob's input, output, or desired computation. Because of blind quantum computing, the privacy is perfect, includes the computation (which is hidden from Steve), does not rely on any computational assumptions, and holds no matter what actions a cheating Steve undertakes.

Alice and Bob only need to prepare single qubits and send them to Steve, who creates an entangled state from all received quantum states. After this initial preparation (where all the quantum communication takes place), Alice and Bob drive the computation, using two-way classical communication to send single-qubit measurement instructions to Steve, depending on previous measurement outcomes. The protocol is characterized by a quantum communication complexity that is $O(m)$, i.e., linear in the input size.

\subsection{Related Work}

An almost complete survey of classical PSI protocols in the honest-but-curious setting can be found in the recent paper by Falk, Noble and Ostrovsky \cite{Falk2019}. In that work, the authors propose a protocol requiring only $O(\sigma m)$ communication between the parties, where $\sigma$ is a security parameter. So far, this is the best classical solution to the PSI problem in the honest-but-curious setting.

Effective protocols for solving the server-aided PSI problem were proposed by Kamara et al. \cite{Kamara2014}. The central idea is to add redundancy to the data sent by Alice and Bob to Steve. Such data are shuffled so that the untrusted server, once executed the set intersection algorithm, cannot easily omit specific items, as the related redundancy is hard to locate. However, there is a case that is not addressed by these protocols: Steve can omit all values from the intersection and simply give back an empty list to Alice and Bob.

Recently, Le et al. \cite{Le2019} proposed a different approach for solving the above cheating potential by Steve without the need of redundancy. The proposed protocols force Steve to prove to Alice and Bob that $z = |A \cap B|$. In the proof, Alice and Bob have to perform a task whose cost is $O(m \log^2 m)$. Moreover, the proposed protocols allow Alice and Bob to compute some arbitrary function over the intersection.   

In the quantum setting, previous work considered the two-party and multi-party PSI scenario \cite{Salman2012,Shi2016,Cheng2017,Maitra2018,Liu2020}. Instead, to the best of our knowledge, the server-aided one has not been considered so far.

\subsection{Outline of the Protocol}

\begin{figure*}[h!]
	\centering
	\includegraphics[width=6cm]{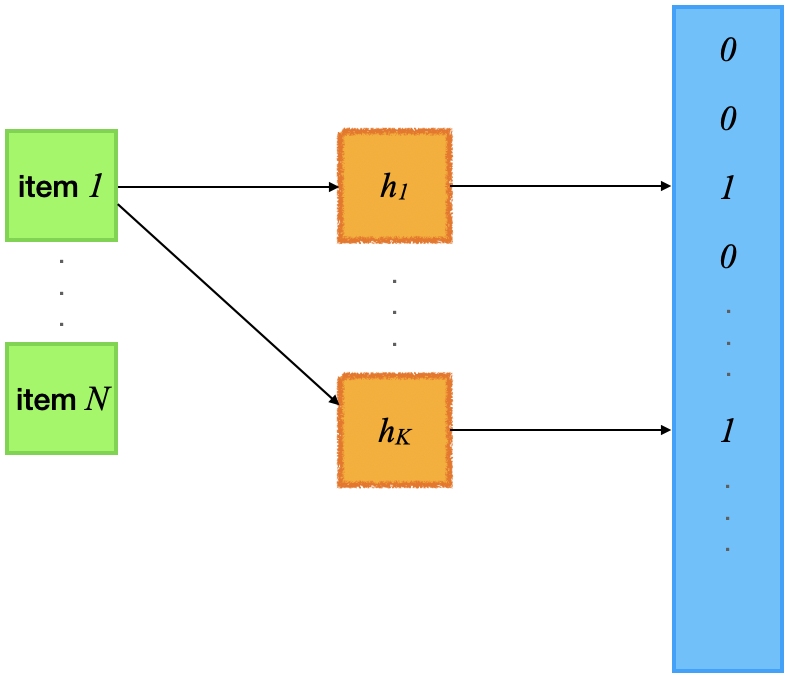}
	\caption{Mapping $N$ items into a Bloom filter, by means of $K$ hash functions.}
	\label{fig:BF}
\end{figure*}

A Bloom filter is a data structure that provides space-efficient storage of sets at the cost of a probability of false positives on membership queries \cite{Broder2004,Amoretti2017}. As illustrated in Fig. \ref{fig:BF}, for each item we set to 1 the $K$ bits of the Bloom filter that correspond to the results of $K$ hash functions computed on the items. The optimal number of hash functions depends on the size $M$ of the Bloom filter and on the maximum number $N$ of items that have to be stored. More precisely, $K_{\text{opt}} = \lfloor \frac{M}{N} \ln 2 \rfloor$. 

In the proposed protocol, Alice and Bob agree on a Bloom filter $F$ of size $\lambda m$, with $\lambda > 1$ sufficiently large to make the probability of false positives negligible. Then, Alice and Bob insert their items in their own Bloom filter instances, respectively $F_\mathcal{A}$ and $F_\mathcal{B}$.
Moreover, Alice stores $D_\mathcal{A}[F_\mathcal{A}(a_i)] \leftarrow a_i$, for all $i = 1,..,m$, in a dictionary $D_\mathcal{A}$.
Similarly, Bob stores $D_\mathcal{B}[F_\mathcal{B}(b_i)] \leftarrow b_i$, for all $i = 1,..,m$, in a dictionary $D_\mathcal{B}$.
The PSI problem reduces to privately computing the bitwise AND of the two Bloom filters.\footnote{It is worth noting that the $\lambda m$-bit Bloom filters allow Alice and Bob to have input sets with a different number of items, provided that such a number is less than or equal to $m$.}

Therefore, using a multiparty delegated quantum computing protocol, Alice and Bob drive Steve into performing a blind quantum computation whose quantum result, which is returned from Steve to Alice and Bob, encodes the classical result \textsf{AND}$_{\text{bitwise}}(F_\mathcal{A},F_\mathcal{B})$. By decrypting the quantum result, Alice and Bob can easily find $\mathcal{A} \cap \mathcal{B}$ by means of their dictionaries. 

If Steve follows the protocol, the output is correct. If Steve is dishonest and deviates from the protocol, anyway he does not learn anything about the inputs of the clients (\textit{security} property) and the computation he is performing (\textit{blindness} property).

\begin{figure*}[t!]
	\centering
	\includegraphics[width=13cm]{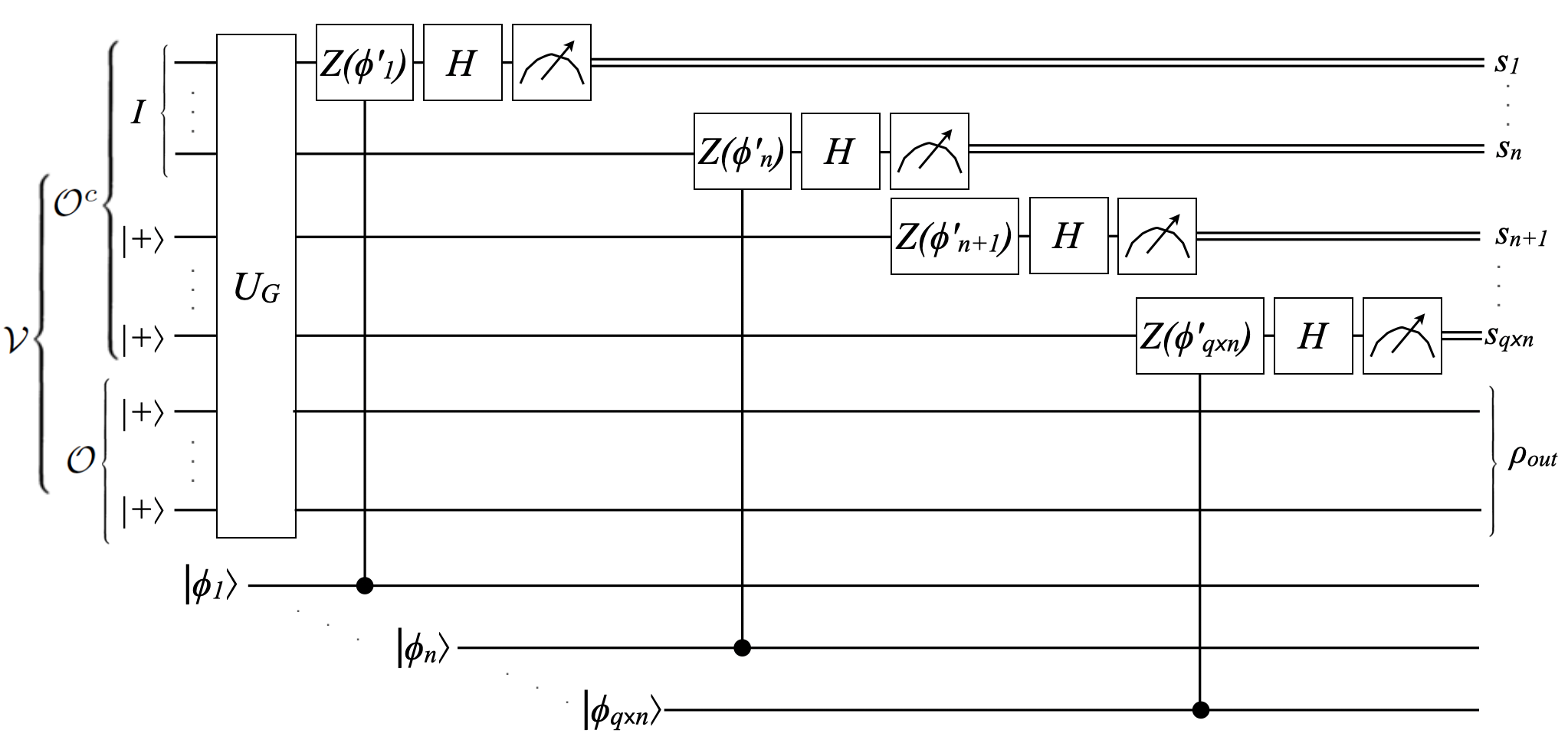}
	\caption{Circuit description of the MBQC model. Here, $U_G$ is a quantum gate that prepares an open graph state.}
	\label{fig:MBQC-circuit}
\end{figure*}

\section{Preliminaries}

\subsection{Measurement-Based Quantum Computing (MBQC)}
In the MBQC model \cite{Raussendorf2001,Raussendorf2003}, a computation is described by a set of measurement angles on an entangled state. 
A circuit description of the MBQC model is provided in Fig. \ref{fig:MBQC-circuit}.
The computation is done in layers, following what is called a \textit{measurement pattern}, which is defined by a finite set of qubits $\mathcal{V}$, a subset of input qubits $\mathcal{I}$, a subset of output qubits $\mathcal{O}$ and a sequence of measurements $\{\phi_j\}$ acting on qubits $\mathcal{O}^c = \mathcal{V} \setminus \mathcal{O}$ (with $\mathcal{I} \subset \mathcal{O}^c$). The outcome of the measurement done at qubit $j$ is denoted as $s_j$. Dependent corrections, used to control nondeterminism, are written as $X_i^{s_j}$ and $Z_i^{s_j}$.

\begin{figure*}[h!]
	\centering
	\includegraphics[width=7cm]{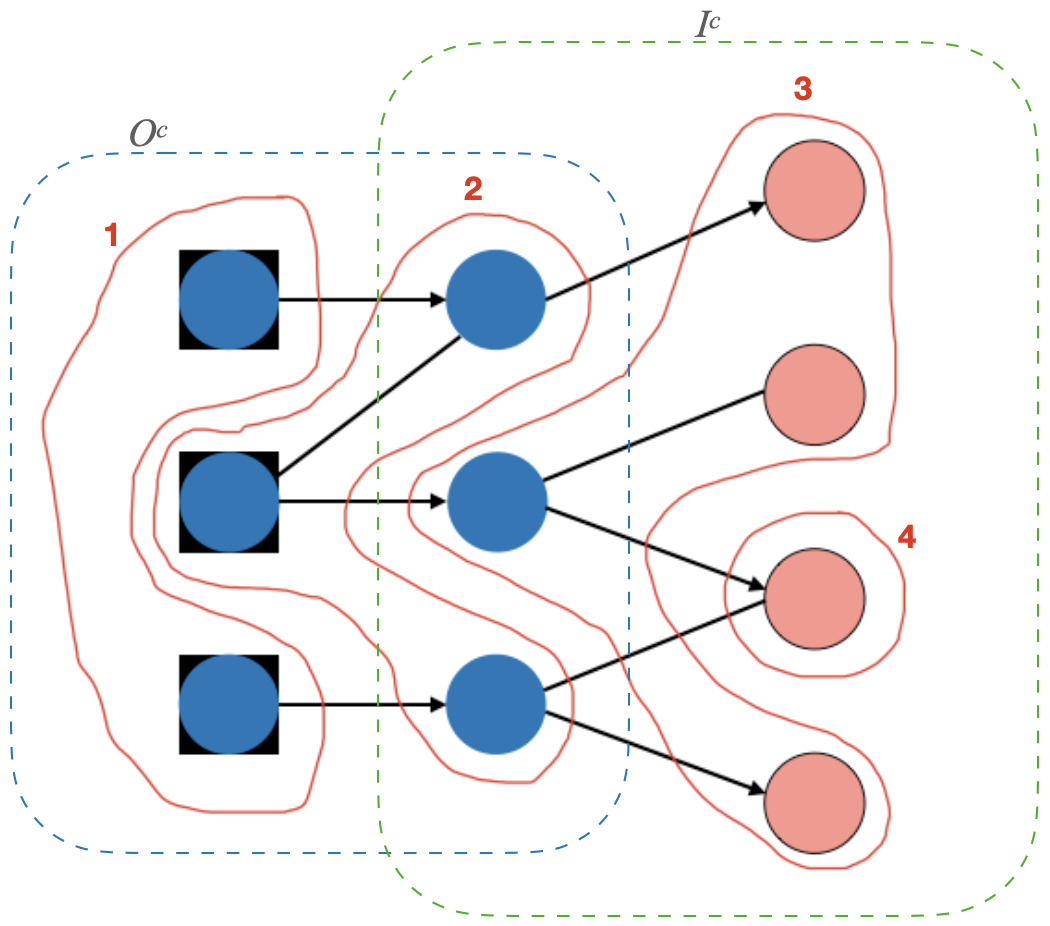}
	\caption{Conceptual representation of an open graph state with flow. The input qubits are depicted as boxed circles, the output qubits as pink circles. All the dark circles (non-output qubits) are measured during the execution of the pattern. The arrows represent the flow function. The 4 partition sets give the partial order on the vertices.}
	\label{fig:OpenGraphState}
\end{figure*}

The qubits in the set $\mathcal{V}$ can be considered as nodes of the underlying undirected graph $G$ of the entangled state. The tuple $(G,\mathcal{I},\mathcal{O})$ defines an \textit{open graph state}, which is prepared by a quantum gate that we denote as $U_G$. In this gate, there are $X$ an $Z$ dependencies and these affect the future measurements. Their placement is dictated by
the \textit{flow} of the graph $G$, which is a map $f: \mathcal{O}^c \rightarrow I^c$, and by a partial order $>$ over the nodes of the graph such that for all $i \in \mathcal{O}^c$:
\begin{itemize}
\item $(i,f(i)) \in G$,
\item $f(i) > i$,
\item for all $k \in \mathcal{N}_G(f(i)) \setminus \{i\}$, we also have $k>i$.
\end{itemize}
Each qubit $j$ is $X$-dependent on qubit $f^{-1}(j)$ and $Z$-dependent on qubits $i$ for which $j$ belongs to the set of neighbors of $f(i)$ in $G$, which is denoted as $\mathcal{N}_G(f(i))$. An open graph state with flow is illustrated in Fig. \ref{fig:OpenGraphState}.

\begin{figure*}[ht!]
	\centering
	\includegraphics[width=15cm]{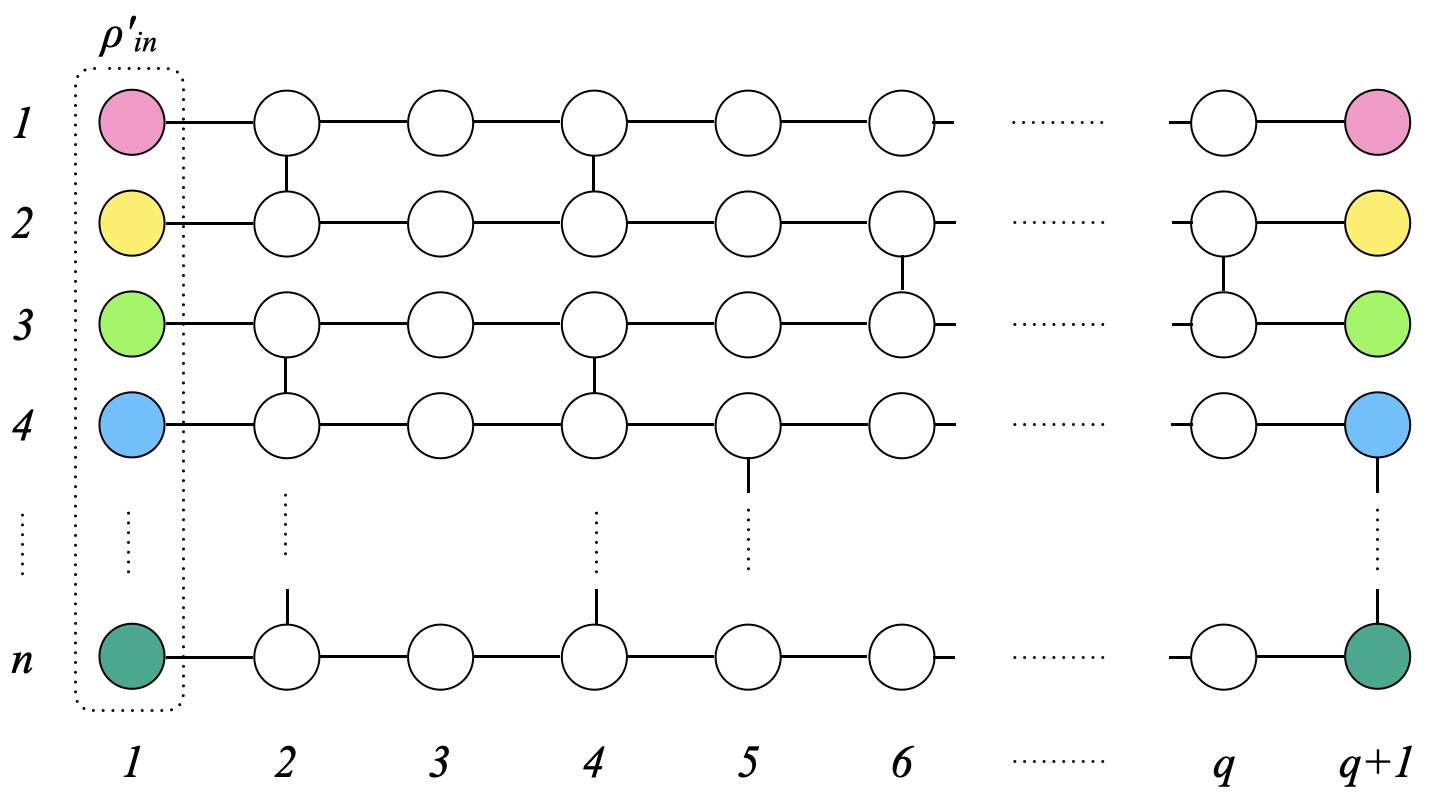}
	\caption{Graph representation of a brickwork state at the server. With reference to the notation used in Fig. \ref{fig:MBQC-circuit}, the total number of qubits is $n \times (q+1)$. The first $n$ qubits at the server are initialized with the encrypted quantum input $\rho'_{in}$. The following $n \times (q-1)$ qubits are prepared to states $\ket{+_{\theta_j}}$, for $j \in O^c \setminus I$. In both cases, protocols that involve all the clients and the server are executed. The last $n$ qubits at the server are prepared in the $\ket{+}$ state. Then, controlled-$Z$ gates are performed between qubits that correspond to vertices joined by an edge, in the graph (the formal construction is reported in \cite{Broadbent2009}). This last step creates the brickwork state at the server. In the implementation, the server does not have all of the qubits of the entire brickwork state in memory at the same time. Indeed, construction and entanglement can be done ``just in time,'' as is standard with measurement-based quantum computing \cite{Chien2015}. Finally, after the execution of the measurement pattern on the brickwork state, the quantum state of the last $n$ qubits will be the output to be returned to the clients.}
	\label{fig:Brickwork}
\end{figure*}

\subsection{Delegated Blind Quantum Computing}

Childs \cite{Childs2005} proposed the first delegated blind quantum computing protocol, with one client and one server, which was quite demanding in terms of quantum resources. In particular, the client was required to control a quantum memory and perform SWAP gates. Later, Arrighi and Salvail \cite{Arrighi2006} introduced a scheme with mechanisms for both verification and blindness for a limited range of functions.

The Universal Blind Quantum Computation (UBQC) scheme by Broadbent, Fitzsimons and E. Kashefi \cite{Broadbent2009} requires the client and the server to exchange only one quantum message, while the rest of the communication is classical. The quantum message sent by the client to the server consists of a tensor product of single-qubit states. Thus, the only quantum capability the client needs is the ability to prepare single-qubit states.

The UBQC protocol is described in terms of the MBQC framework. 
More precisely, UBQC can be considered as the distributed version of an MBQC computation. In this context, any quantum computation (represented by a unitary operator $U$) is given as a measurement pattern on a \emph{brickwork} state, which is an entangled state of $n \times (q+1)$ qubits. 

In the \emph{preparation phase}, the client prepares $n \times (q+1)$ quantum states and sends them to the server, which entangles them for creating the brickwork state. Note that this process unavoidably reveals upper bounds on the length of the input and depth of the computation. However, due to universality of the brickwork state, it does not reveal any additional information on the client's computation.

The client has in mind a unitary operator $U$ that is implemented with a measurement pattern on the brickwork state. This pattern could have been designed either directly in MBQC or from a circuit construction. Note that it is assumed that the client's input to the computation is built into $U$. In other words, the client wishes to compute $U\ket{0}$. In the \emph{computation phase}, the client transmits (classical) measurement instructions to the server. The classical outcomes of the measurements are communicated by the server to the client, whose choice of the angles in future  rounds will depend  on these values. The protocol is blind as the client's quantum states and classical messages are astutely chosen so that, no matter what the server does, it cannot infer anything about the client's measurement pattern. At the end, the server returns the final qubits to the client.

To further extend the idea of computing over encrypted data, a multiparty delegated quantum computation protocol in the MBQC framework was proposed by Kashefi and Pappa \cite{Kashefi2017}. Also this protocol consists of a preparation phase where all the quantum communication takes place, followed by a computation phase where the communication is purely classical. During the first stage, each client one-time pads its quantum input and sends it to the server. In this way, the private data of the clients remain secret during the protocol. At the end of this stage, the server entangles the received quantum states in order to produce the brickwork state (Fig. \ref{fig:Brickwork}). In the second stage, the clients need to securely communicate between them and with the server, in order to jointly compute the measurement angles of the qubits in the different layers of computation. This procedure is purely classical, and based on a \textit{Verifiable Secret Sharing (VSS)} scheme \cite{Chor1985} and a \textit{computation oracle} \cite{Canetti2001,Ishai2008,Unruh2010}. The resulting measurement pattern does not reveal the corresponding unitary operator $U$ to the server. At the end of the MBQC process, each client receives its quantum output, consisting of qubits that are naturally encrypted due to the randomness from previous measurements that propagated during the computation. The decryption of the quantum output is based on the classical secret shares of all clients.

\section{Main Protocol}
\label{sec:protocol}
 
From now on, we denote Alice and Bob (the two clients) as $C_1$ and $C_2$. We want that the private data of each client (i.e., its quantum input and output) remains secret during the protocol. Moreover, we want that the measured angles are not known to the server (i.e., Steve), but are secret-shared with the clients (using a VSS scheme). The quantum input provided by the two clients contributes to the preparation of the quantum state $\rho'_{in}$ at the server. The measurement angles for qubits $j \in \mathcal{O}^c$ are denoted as $\{\phi_j\}$.  
 
\subsection{Preparation Phase}

The quantum state $\rho'_{in}$ at the server is prepared as follows. For $j \in \mathcal{I}$ (such that $n$ qubits of the server are affected):
\begin{enumerate}
\item The client that owns the $j$th qubit applies $X^{c_j}Z(\theta_j^j)$ to its qubit (i.e., performs a one-time padding) and sends the quantum state to the server. The values $c_j \in \{0,1\}$ and $\theta_j^j \in \{l\pi/4\}_{l=0}^7$ are randomly picked, and secret-shared with the other client.
\item The other client runs Protocol 1 with the server. If the client passes the test, the server at the end has the state $\ket{+_{\theta_j^k}} = \frac{1}{\sqrt{2}}(\ket{0} + e^{i\theta_j^k}\ket{1})$, where $k \in \{1,2\}$ identifies the client.
\item The server runs Protocol 2 and announces the outcome $t_j$.
\end{enumerate}
At this point the server has the state 
\begin{equation}
\rho'_{in} = (X^{c_1}Z(\theta_1) \otimes ... \otimes X^{c_n}Z(\theta_n))\rho_{in}
\end{equation}
where 
\begin{equation}
\theta_j = \theta^j_j + \sum_{k=1,k \neq j}^n (-1)^{\oplus_{i=k}^n t^i_j +c_j} \theta^k_j
\end{equation}

\noindent
Then, for $j \in \mathcal{O}^c \setminus \mathcal{I}$ (such that $n \times (q-1)$ qubits of the server are affected):
\begin{enumerate}
\item Both clients run Protocol 1 with the server. If the test is passed by both clients, the server at the end has two states $\ket{+_{\theta_j^k}} = \frac{1}{\sqrt{2}}(\ket{0} + e^{i\theta_j^k}\ket{1})$, where $k \in \{1,2\}$ identifies the client.
\item The server runs Protocol 3 getting the outcome $t_j$, and ends up with the state $\ket{+_{\theta_j}}$, where
\begin{equation}
\label{eq:theta_j}
\theta_j = \theta_j^2 + (-1)^{t_j}\theta_j^1 
\end{equation}
\end{enumerate}

\noindent
At this point, for $j \in \mathcal{O}$, the server prepares $\ket{+}$ states (such that $n$ qubits of the server are affected).
Finally, the server entangles all the $n \times (q+1)$ qubits to a brickwork state by applying $CZ$ gates.

\begin{framed}
\noindent
\textbf{Protocol 1} (To enforce honest behavior for client $C_k$)
\begin{enumerate}
\item Client $C_k$ sends $L$ quantum states $\ket{+_{\theta_l^k}} = \frac{1}{\sqrt{2}}(\ket{0} + e^{i\theta_l^k}\ket{1})$ to the server, and secret-shares the values $\{\theta_l^k\}_{l=1,..,L}$ with the other client, using a VSS scheme. 
\item The server requests the shared values from the clients, for all but one qubit, and measures in the bases reconstructed by means of the shared values. If the bases agree with the measurements, then the remaining state is correctly formed in relation to the remaining shared angle, with high probability.
\end{enumerate}
\end{framed}

\begin{framed}
\noindent
\textbf{Protocol 2} (State preparation for $j \in \mathcal{I}$)
The server stores the quantum states received by the two clients, namely $X^{c_j}Z(\theta_j^j)\ket{\psi_j}$ and $\ket{+_{\theta_j^k}}$, to distinct registers. Then, a CNOT is applied, and finally the second qubit is measured, producing the outcome $t_j$ and the quantum state $X^{c_j}Z(\theta_j)\ket{\psi_j}$, where $\theta_j = \theta_j^j + (-1)^{t_j + c_j}\theta_j^k$.  
\begin{center}
	\centering
	\includegraphics[width=7cm]{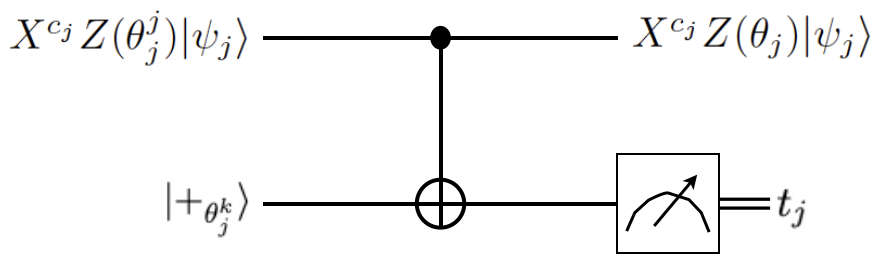}
	\label{fig:Protocol2}
\end{center}
\end{framed}

\begin{framed}
\noindent
\textbf{Protocol 3} (State preparation for $j \in \mathcal{O}^c \setminus \mathcal{I}$)
The server stores the quantum states received by the two clients, namely $\ket{+_{\theta_j^1}}$ and $\ket{+_{\theta_j^2}}$, to distinct registers. Then, a CNOT is applied, and finally the first qubit is measured, producing the outcome $t_j$ and the quantum state $\ket{+_{\theta_j}}$.
\begin{center}
	\centering
	\includegraphics[width=5cm]{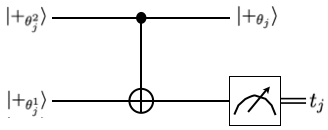}
	\label{fig:Protocol3}
\end{center}
\end{framed}

\subsection{Computation Phase}

In this phase, the clients drive the measurement process at the server. 
We remark that, in the following description, $s_j^X$ denotes the sum of the values $s_i$, where $i$ is each node of the graph (representing the entangled state) that has an $X$-dependency with node $j$. Similarly, $s_j^Z$ denotes the sum of the values $s_i$, where $i$ is each node that has a $Z$-dependency with node $j$.

\noindent
For $j \in \mathcal{O}^c$ (the $n \times q$ non-output qubits of the server):
\begin{enumerate}
\item Both clients choose random $r_j^k \in \{0,1\}$, which they secret-share with each other. Then, with the help of a computation oracle, they compute the measurement angle of qubit $j$:
\begin{equation}
\delta_j = \phi'_j + \pi r_j + \theta_j
\end{equation}
where undefined values are equal to zero, or otherwise:
\begin{itemize}
\item $\phi'_j = (-1)^{c_j + s_j^X}\phi_j + s_j^Z\pi + c_{f^{-1}(j)}\pi$
\item $\phi_j$ is the ``plain'' measurement angle for the $j$th qubit in the brickwork state, 
\item $s_j^X = \sum_{i \text{ X-dep. on } j} s_i$, 
\item $s_j^Z = \sum_{i \text{ Z-dep. on } j} s_i$, 
\item $s_i = m_i \oplus r_i$, for $i \leq j$,
\item $r_j = r_j^1 + r_j^2$. 
\end{itemize}
\item The server receives $\delta_j$ and measures qubit $j$ in the $\{\ket{+_{\delta_j}},\ket{-_{\delta_j}}\}$ basis, getting $m_j$ as result, which is then announced by the server to the clients. 
\end{enumerate}
Then, for $j \in \mathcal{O}$ (the last $n$ qubits at the server), the server sends the quantum state to the corresponding client, which applies $Z^{s_j^Z}X^{s_j^X}$ to retrieve the actual quantum output. 

A straightforward way to implement the bitwise AND function between the inputs of the clients is to implement $\lambda m$ parallel Toffoli gates in the brickwork state.\footnote{For each Toffoli gate, there are two input qubit states and one ancilla $\ket{0}$ state that must be provided by one of the clients.}
However, the encrypted quantum output of each Toffoli gate cannot be produced in two copies (one for each client), because of the no-cloning theorem. To solve this issue, half of the quantum output is sent to $C_1$, the other half to $C_2$. Therefore, further secure classical interaction between the clients is needed, so that they both end up with \textsf{AND}$_{\text{bitwise}}(F_\mathcal{A},F_\mathcal{B})$.

\section{Analysis of the Protocol}

\subsection{Communication Complexity}
To implement $\lambda m$ parallel Toffoli gates in the brickwork state, using the strategy proposed by Chien et al. \cite{Chien2015} we end up with $q+1=57$ layers, each layer having $n=3 \lambda m$ qubits. The total number of transmitted qubits is then $n \times (q+1) = 171 \lambda m$. For each transmitted quantum state, there is an overhead due to Protocols 1, 2 and 3, that can be expressed as a constant factor. We may conclude that, in general, the quantum communication complexity of the protocol is $O(m)$.

\subsection{Correctness, Security and Blindness}
The \textit{correctness} of the proposed protocol comes from the correctness of the individual circuits implementing Protocols 2 and 3. A detailed proof can be derived from the proof of Theorem 1 in \cite{Kashefi2017}, asserting the correctness of the general MBQC-based multiparty delegated quantum computation protocol.

The \textit{security} property of the protocol derives from the fact that the clients never share their sets with each other. 
The proposed protocol is also secure against a malicious server. This is true in general for the MBQC-based multiparty delegated quantum computation protocol \cite{Kashefi2017}. The proof is based on the fact that the protocol emulates an ideal multiparty delegated quantum computation resource that does not give the server any access to the clients' input. 

At the same time, the protocol has the \textit{blindness} property, meaning that the server does not know what computation it is doing, provided that the measurement angles $\{\phi_j\}$ remain hidden from the server.
 
\section{Leveraging the Quantum Internet}

\begin{figure}[ht!]
	\centering
	\includegraphics[width=14cm]{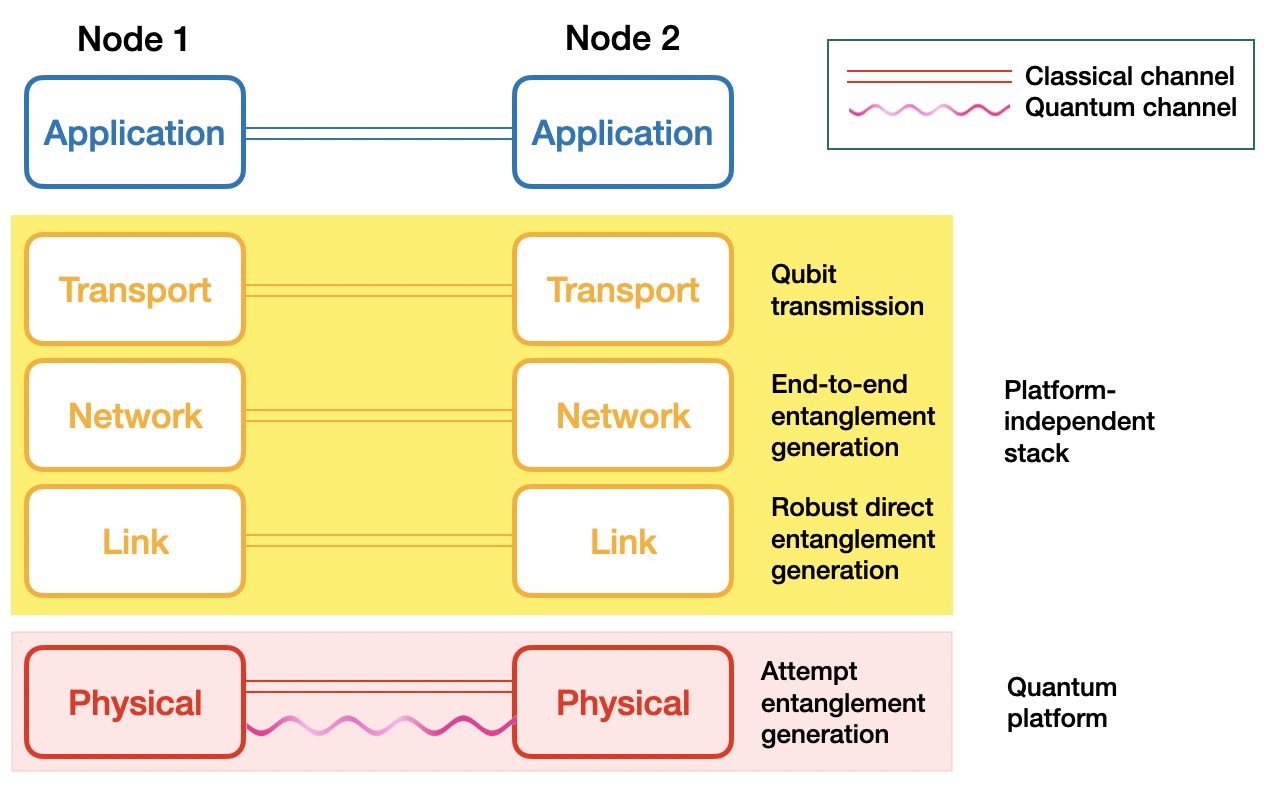}
	\caption{Quantum network stack architecture \cite{Dahlberg2019}. A link layer and a physical layer protocol have been experimentally demonstrated by Pompili et al. \cite{Pompili2021}.}
	\label{fig:QuantumNetworkStack}
\end{figure}

In the proposed protocol, Alice and Bob send qubits to Steve in the preparation phase, while Steve sends qubits to Alice and Bob at the end of the computation phase.
In order to move qubits between any two parties over long distances, \textit{quantum state teleportation} is preferred to quantum communication, whose fidelity decreases exponentially with the channel length, due to loss \cite{Rohde2021}. Quantum teleportation requires \textit{end-to-end entanglement generation}, i.e., probably the most important general-purpose service in the future Quantum Internet \cite{Wehner2018,Cacciapuoti2020,Nguyen2017,Amoretti2020,VanMeter2016,Ferrari2021}. In Figure \ref{fig:QuantumNetworkStack}, an example of quantum network stack architecture, inspired by the TCP/IP one, is presented \cite{Dahlberg2019,Pompili2021}. 

\section{Conclusion and Future Work}

In this work we have proposed a protocol that solves the server-aided PSI problem using delegated blind quantum computing. The protocol is correct, secure and blind against a malicious server. Moreover, it is characterized by a quantum communication complexity that is linear in the input size.

Regarding future work, we plan to study alternative approaches to the implementation of the bitwise AND function at the server. Moreover, we are interested in exploring different strategies with respect to the Bloom filter's one we have adopted in this work, to efficiently map the clients' input to the server. 
An interesting future direction is investigating the possibility to implement a variant of the proposed protocol where clients are fully classical.
In a recent work, Aaronson, Cojocaru, Gheorghiu and Kashefi \cite{Aaronson2019} suggested that there is no scheme for blind quantum computing that is information-theoretically secure and that requires only classical communication between client and server. On the other hand, there are interesting proposals for fully-classical client protocols that achieve more restricted levels of security \cite{Mantri2017,Mahadev2018,Cojocaru2021}.

\section*{Acknowledgements}
The author would like to thank Anna Pappa for helpful discussions on the MBQC-based multiparty delegated quantum computation protocol.


\begin{thebibliography}{00}

\bibitem{Brickell2007}
J. Brickell, D. E. Porter, V. Shmatikov, E. Witchel,
\emph{Privacy-Preserving Remote Diagnostics},
Proc. of the 14th ACM conference on Computer and Communications Security (2007)

\bibitem{He2017}
X. He, A. Machanavajjhala, C. Flynn, D. Srivastava,
\emph{Composing Differential Privacy and Secure Computation: A case study on scaling private record linkage},
Proc. of the 24th ACM conference on Computer and Communications Security (2017)

\bibitem{Demmler2018}
D. Demmler, P. Rindal, M. Rosulek, N. Trieu,
\emph{PIR-PSI: Scaling Private Contact Discovery},
Proc. on Privacy Enhancing Technologies, vol. 2018, no. 4 (2018)

\bibitem{Yanai2020}
A. Yanai,
\emph{Private Set Intersection},
\url{https://decentralizedthoughts.github.io/2020-03-29-private-set-intersection-a-soft-introduction/} (2020)

\bibitem{Kashefi2017}
E. Kashefi, A. Pappa,
\emph{Multiparty Delegated Quantum Computing},
Cryptography, vol. 1, no. 12 (2017)

\bibitem{Falk2019}
B. Hemenway Falk, D. Noble, R. Ostrovsky,
\emph{Private Set Intersection with Linear Communication from General Assumptions},
Proc. of the 18th ACM Workshop on Privacy in the Electronic Society (2019)

\bibitem{Kamara2014}
S. Kamara, P. Mohassel, M. Raykova, S. Sadeghian,
\emph{Scaling Private Set Intersection to Billion-Element Sets}, 
Proc. of the International Conference on Financial Cryptography and Data Security (2014)

\bibitem{Le2019}
P. H. Le, S. Ranellucci, S. Dov Gordon,
\emph{Two-party Private Set Intersection with an Untrusted Third Party},
Proc. of the 2019 ACM SIGSAC Conference on Computer and Communications (2019)

\bibitem{Cheng2017}
X. Cheng, R. Guo, Y. Chen,
\emph{Cryptanalysis and improvement of a quantum private set intersection protocol},
Quantum Information Processing, vol. 16, no. 37 (2017)

\bibitem{Liu2020}
B. Liu, M. Zhang, R. Shi,
\emph{Quantum Secure Multi-party Private Set Intersection Cardinality},
International Journal of Theoretical Physics vol. 59, pp. 1992--2007 (2020)

\bibitem{Maitra2018}
A. Maitra,
\emph{Quantum secure two-party computation for set intersection with rational players},
Quantum Information Processing, vol. 17, no. 197 (2018) 

\bibitem{Salman2012}
T. Salman and Y. Baram,
\emph{Quantum Set Intersection and its Application to Associative Memory},
Journal of Machine Learning Research 13 (2012) 

\bibitem{Shi2016}
R. Shi, Y. Mu, H. Zhong, J. Cui, S. Zhang,
\emph{An efficient quantum scheme for Private Set Intersection}, 
Quantum Information Processing, vol. 15, no. 1 (2016)

\bibitem{Broder2004}
A. Broder, M. Mitzenmacher, 
\emph{Network applications of Bloom filters: A survey}, 
Internet Math., vol. 1, no. 4, pp. 485–509 (2004)

\bibitem{Amoretti2017}
M. Amoretti, O. Alphand, G. Ferrari, F. Rousseau and A. Duda, 
\emph{DINAS: A Lightweight and Efficient Distributed Naming Service for All-IP Wireless Sensor Networks}, 
IEEE Internet of Things Journal, vol. 4, no. 3, pp. 670-684 (2017)

\bibitem{Raussendorf2001}
R. Raussendorf, H. Briegel,
\emph{A one-way quantum computer},
Phys. Rev. Lett., vol. 86, pp. 5188--5191 (2001)

\bibitem{Raussendorf2003}
R. Raussendorf, D. Browne, H. Briegel,
\emph{Measurement-based quantum computation with cluster states},
Phys. Rev. A, vol. 68, pp. 022312 (2003)

\bibitem{Aaronson2019}
S. Aaronson, A. Cojocaru, A. Gheorghiu, E. Kashefi,
\emph{Complexity-theoretic limitations on blind delegated quantum computation},
Proc. of the 46th International Colloquium on Automata, Languages, and Programming (2019)

\bibitem{Childs2005}
A. Childs,
\emph{Secure Assisted Quantum Computation},
Quantum Information \& Computation, vol. 5, no. 6, pp. 456--466 (2005)

\bibitem{Arrighi2006}
P. Arrighi, L. Salvail,
\emph{Blind Quantum Computation},
International Journal of Quantum Information, vol. 4, no. 5, pp. 883--898 (2006) 

\bibitem{Broadbent2009}
A. Broadbent, J. Fitzsimons, E. Kashefi,
\emph{Universal Blind Quantum Computation},
Proc. of the 50th Annual Symposium on Foundations of Computer Science (2009)

\bibitem{Chien2015}
C.-H. Chien, E. van Meter, S.-Y. Kuo,
\emph{Fault-Tolerant Operations for Universal Blind Quantum Computation}
ACM Journal on Emerging Technologies in Computing Systems, vol. 12, no. 1, 2015.

\bibitem{Chor1985}
B. Chor, S. Goldwasser, S. Micali, B. Awerbuch, 
\emph{Verifiable secret sharing and achieving simultaneity in the presence of faults}, 
26th Annual Symposium on Foundations of Computer Science, pp. 383-395 (1985)

\bibitem{Canetti2001}
R. Canetti,
\emph{Universally composable security: a new paradigm for cryptographic protocols},
Proc. of the 42nd IEEE Symposium on Foundations of Computer Science (2001)	

\bibitem{Ishai2008}
Y. Ishai, M. Prabhakaran, A. Sahai,
\emph{Founding Cryptography on Oblivious Transfer -- Efficiently}
Proc. of CRYPTO (2008)

\bibitem{Unruh2010}
D. Unruh,
\emph{Universally Composable Quantum Multi-party Computation},
Proc. of EUROCRYPT (2010) 

\bibitem{Rohde2021}
P. P. Rohde,
\emph{The Quantum Internet},
Cambridge University Press (2021)

\bibitem{Wehner2018}
S. Wehner, D. Elkouss, R. Hanson, 
\emph{Quantum Internet: a Vision for the road ahead}, 
Science, 362, 6412 (2018)

\bibitem{Cacciapuoti2020}
A. S. Cacciapuoti, M. Caleffi, F. Tafuri, F. S. Cataliotti, S. Gherardini and G. Bianchi,
\emph{Quantum Internet: Networking Challenges in Distributed Quantum Computing}, 
IEEE Network, vol. 34, no. 1, pp. 137-143 (2020)

\bibitem{Nguyen2017}
H. V. Nguyen, Z. Babar, D. Alanis, P. Botsinis, D. Chandra, M. A. Mohd Izhar, S. X. Ng and L. Hanzo,
\emph{Towards the Quantum Internet: Generalised Quantum Network Coding for Large-Scale Quantum Communication Networks}, 
IEEE Access, vol. 5, pp. 17288-17308 (2017)

\bibitem{Amoretti2020}
M. Amoretti and S. Carretta,
\emph{Entanglement verification in quantum networks with tampered nodes},
IEEE Journal on Selected Areas in Communications, vol. 38, no. 3, pp. 598-604 (2020)

\bibitem{VanMeter2016}
R. Van Meter and S. J. Devitt, 
\emph{The Path to Scalable Distributed Quantum Computing}, 
Computer, vol. 49, no. 9, pp. 31-42 (2016)

\bibitem{Ferrari2021}
D. Ferrari, A. S. Cacciapuoti, M. Amoretti and M. Caleffi, 
\emph{Compiler Design for Distributed Quantum Computing}, 
IEEE Transactions on Quantum Engineering, vol. 2, pp. 1-20, art no. 4100720 (2021)

\bibitem{Dahlberg2019}
A. Dahlberg, M. Skrypczyk, T. Coopmans, L. Wubben, F. Rozpedek, M. Pompili, A. Stolk, P. Pawelczak, R. Knegjens, J. de Oliveira Filho, R. Hanson, S. Wehner,
\emph{A link layer protocol for quantum networks},
Proc. of ACM SIGCOMM, pp.159-173 (2019)

\bibitem{Pompili2021}
M. Pompili, C. Delle Donne, I. te Raa, B. van der Vecht, M. Skrypczyk, G. Ferreira, L. de Kluijver, A. J. Stolk, S. L. N. Hermans, P. Pawelczak, W. Kozlowski, R. Hanson, S. Wehner, 
\emph{Experimental demonstration of entanglement delivery using a quantum network stack},
arXiv:2111.11332 (2021)

\bibitem{Mantri2017}
A. Mantri, T. F. Demarie, N. C. Menicucci, J. F. Fitzsimons, 
\emph{Flow Ambiguity: A Path Towards Classically Driven Blind Quantum Computation},
Physical Review X, vol. 7, no. 3, pp. 031004 (2017)

\bibitem{Mahadev2018}
U. Mahadev,
\emph{Classical Verification of Quantum Computations},
Proc. of IEEE FOCS, pp.259-267 (2018)

\bibitem{Cojocaru2021}
A. Cojocaru, L. Colisson, E. Kashefi and P. Wallden, 
\emph{On the Possibility of Classical Client Blind Quantum Computing},
Cryptography, vol. 5, no. 1, art no. 3 (2021)



\end{thebibliography}
\end{document}